\documentclass[a4paper,english,aps,prl,nobibnotes,reprint,superscriptaddress,showpacs,showkeys,longbibliography,floatfix]{revtex4-2}

\usepackage[utf8]{inputenc}
\usepackage[T1]{fontenc}
\usepackage{lmodern}
\input{glyphtounicode}
\usepackage{babel}
\usepackage{mathtools}
\usepackage{amsmath}
\usepackage{amssymb}
\usepackage{graphicx}
\usepackage{csquotes}
\usepackage{float}
\usepackage[colorlinks=true, linkcolor=red, citecolor=blue, urlcolor=blue]{hyperref}
\usepackage[usenames,dvipsnames]{color}
\usepackage{bbold}
\usepackage{soul}
\usepackage{footmisc}

\usepackage{siunitx}
\usepackage{physics} 

\renewcommand{\vec}[1]{\boldsymbol{#1}} 

\definecolor{mygreen}{rgb}{0,0.5,0}
\definecolor{myblue}{rgb}{0,0,0.75}
\definecolor{mymagenta}{cmyk}{0,1,0,0.12}

\usepackage{xcolor}
\usepackage{xspace}

\newcommand{\DMM}{[1-${}^{13}\mathrm{C}$,$\mathrm{d}_6$]-dimethyl maleate\xspace}
\newcommand{\DMAD}{[1-${}^{13}\mathrm{C}$,$\mathrm{d}_6$]-dimethyl acetylenedicarboxylate\xspace}

\newcommand{\Cth}{${}^{13}\mathrm{C}$\xspace}
\newcommand{\Hone}{${}^{1}\mathrm{H}$\xspace}

\newcommand{\Cat}{[Rh(dppb)(COD)]BF$_4$\xspace}

\begin{document}
	
\title{Robust Parahydrogen-Induced Polarization at High Concentrations}

\author{L. Dagys}
\affiliation{NVision Imaging Technologies GmbH, Wolfgang-Paul Stra{\ss}e 2, 89081 Ulm, Germany}
\author{M. C. Korzeczek}
\affiliation{Institut f\"{u}r Theoretische Physik \& IQST, Albert-Einstein Allee 11, Universit\"{a}t Ulm, D-89081 Ulm, Germany}
\author{A. J. Parker}
\affiliation{NVision Imaging Technologies GmbH, Wolfgang-Paul Stra{\ss}e 2, 89081 Ulm, Germany}
\author{J. Eills}
\affiliation{Institute of Bioengineering of Catalonia,  08028 Barcelona, Spain}
\author{J. W. Blanchard} 
\affiliation{Quantum Technology Center, University of Maryland, MD 20742, United States}
\author{C. Bengs}
\affiliation{University of Southampton, Southampton, United Kingdom, SO171BJ}
\author{M. H. Levitt}
\affiliation{University of Southampton, Southampton, United Kingdom, SO171BJ}
\author{S. Knecht}
\affiliation{NVision Imaging Technologies GmbH, Wolfgang-Paul Stra{\ss}e 2, 89081 Ulm, Germany}
\author{I. Schwartz}
\email{ilai@nvision-imaging.com}
\affiliation{NVision Imaging Technologies GmbH, Wolfgang-Paul Stra{\ss}e 2, 89081 Ulm, Germany}

\author{M. B. Plenio}
\email{martin.plenio@uni-ulm.de}
\affiliation{Institut f\"{u}r Theoretische Physik \& IQST, Albert-Einstein Allee 11, Universit\"{a}t Ulm, D-89081 Ulm, Germany}

\begin{abstract}
Parahydrogen-Induced Polarization (PHIP) is a potent technique for generating target 
molecules with high nuclear spin polarization. The PHIP process involves 
a chemical reaction between parahydrogen and a target molecule, followed by the 
transformation of nuclear singlet spin order into magnetization of a designated nucleus 
through magnetic field manipulations. Although the singlet-to-magnetization polarization
transfer process works effectively at moderate concentrations, it is observed to become
much less efficient at high molar polarization, defined as the product of polarization and 
concentration. This strong dependence on the molar polarization is attributed to interference 
from the field produced by the sample's magnetization during polarization transfer, which 
leads to complex dynamics and can severely impact the scalability of the technique. We 
address this challenge with a  pulse sequence that negates the influence of the distant 
dipolar field, while simultaneously achieving singlet-to-magnetization polarization transfer 
to the desired target spins, free from restrictions
on the molar polarization. 
\end{abstract}
	
\maketitle
\date{\today }

{\em Introduction --}
Nuclear Magnetic Resonance (NMR), one of the most widespread spectroscopic techniques with a 
broad range of applications, extending from chemical analysis and drug discovery to medical 
imaging, is intrinsically limited by its low sensitivity. This limitation is rooted in the 
weak nuclear spin polarization in thermal equilibrium, typically amounting to a few parts per 
million. 
Thermal-equilibrium polarization and detection can be improved by increasing magnetic 
field strength which may not be easily achievable.
A promising alternative to address the sensitivity challenge involves hyperpolarization methods, 
which can enhance nuclear spin polarization by orders of magnitude compared to the level at 
thermal equilibrium \cite{eills_spin_2023,Ardenkjær-Larsen_increase_2003,maly_dynamic_2008,eichhorn_hyperpolarized_2022,navon_enhancement_1996,marco-rius_quantitation_2014,Ledbetter_nonlinear_2002,bowers_transformation_1986,bowers_parahydrogen_1987,natterer_parahydrogen_1997,reineri_parahydrogen_2015,buntkowsky_recent_2022,knecht_rapid_2021,goldman_metabolic_2006,korchak_spontaneous_2021,marshall_radio-frequency_2023,Dagys_deuteron_2022,D1CP04653E}.

Parahydrogen-Induced Polarization (PHIP) \cite{bowers_transformation_1986,bowers_parahydrogen_1987,natterer_parahydrogen_1997,reineri_parahydrogen_2015,buntkowsky_recent_2022,knecht_rapid_2021,goldman_metabolic_2006,korchak_spontaneous_2021,marshall_radio-frequency_2023,Dagys_deuteron_2022,D1CP04653E}
is a hyperpolarization method that offers a high level of polarization 
and fast throughput of polarized samples. PHIP involves an irreversible hydrogenation reaction between 
a substrate and para-enriched hydrogen (parahydrogen) gas which is used to embed the nuclear singlet
order of parahydrogen in newly formed product molecules. Upon completion of the reaction, the singlet
order is then transformed into observable magnetization using a variety of methods, e.g., coherence transfer
by NMR pulse sequences or adiabatic transfer schemes \cite{johannesson2004transfer,goldman2005hyperpolarization,korchak_spontaneous_2021,marshall_radio-frequency_2023,Dagys_deuteron_2022,D1CP04653E,devience_preparation_2013,eills2019polarization,Pravdivtsev2016RobustCO,RODIN201814,BENGS2020106850,pileio_storage_2010,eills_singlet_2017}.
Consequently, PHIP can generate samples 
with molar polarization, defined as the product of the spin polarization and the concentration of target 
nuclei, reaching reported values of around $50$--$100\,$mM molar polarization for \Cth in fumarate \cite{knecht_rapid_2021}. 

The NMR signal is proportional to molar polarization, which is a better figure of merit than polarization alone for many applications such as metabolic imaging or fundamental physics experiments, for which high polarization alone is insufficient and high target concentrations are also desired~\cite{knecht_rapid_2021,goldman_metabolic_2006,blanchard2021towards}.
Additionally, it may unlock applications that inherently benefit from high sample magnetization, 
such as microscale NMR \cite{Schwartz2019,eills2022synergies} or the nuclear Overhauser effect methods in liquid samples
\cite{navon_enhancement_1996,marco-rius_quantitation_2014,korchak_spontaneous_2021,eichhorn_hyperpolarized_2022}. 
Hence, it is pertinent to inquire to what extent achievable molar polarization can be increased.

In this context it is important to note that high molar polarization can introduce adverse effects.
For example, a sample of \Hone water only yields about $3\,$mM of \Hone  molar polarization at 
9~T magnetic field and room temperature (111\,M \Hone concentration at 0.003\% polarization), but this is sufficient intrinsic 
magnetization to act back on the sample itself. After rf excitation, such magnetization in a tuned rf coil induces a current 
that generates an additional transverse field that rotates sample magnetization out of phase and 
causes radiation damping~\cite{broekaert_suppression_1995,Krishnana2013}.
This typically leads to line broadening, phase distortions and other effects often associated 
with ${}^{1}\mathrm{H}$- and ${}^{19}\mathrm{F}$-rich samples. 

A less pronounced phenomenon does 
not require coupling to a tuned coil, and emerges from the (small) nuclear spin contribution to the magnetic flux density of 
the sample~\cite{edzes_the_1990,warren_generation_1993,levitt_demagnetization_1996,pelupessy_transfer_2022,deville_NMR_1979}.
A cylindrical 100~mM sample of \Hone spins at 50\% polarization (50~mM molar polarization) can generate a magnetic flux density of 180~nT corresponding to an 8\,Hz resonance shift, while the previous example of water placed in a 9~T magnetic field would result in a $0.5\,$Hz shift~\cite{edzes_the_1990,warren_generation_1993,levitt_demagnetization_1996}. The backaction of these internal fields is 
known to induce chaotic dynamics even in highly symmetric samples with 
uniform initial polarization distribution, as even minute inhomogeneities can be amplified
rapidly~\cite{Ledbetter_nonlinear_2002,jeener_dynamical_1999,jeener_dynamical_2002,levitt_demagnetization_1996,warren_generation_1993}.

\begin{figure}[h]
    \includegraphics[width=0.45\textwidth]{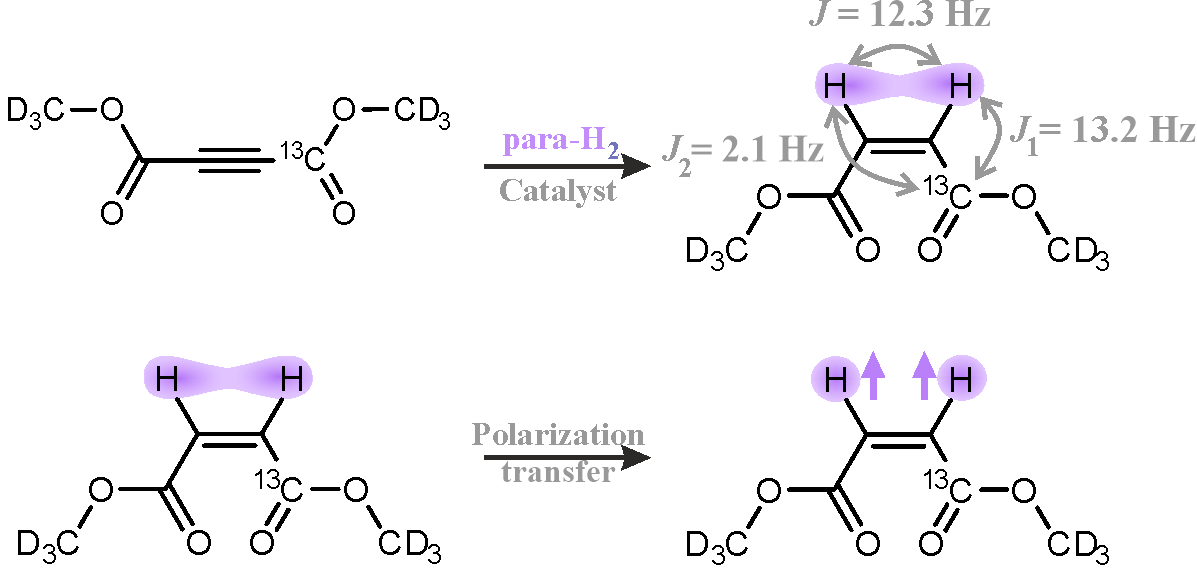}
    \centering
    \caption{Hyperpolarization of \DMM using PHIP. Top - hydrogenation reaction of \DMAD using parahydrogen yields \DMM with two protons in a nuclear singlet state. $J$-couplings are indicated and taken from \cite{BENGS2020106850}. Deuterons and their couplings are ignored in the theory and simulations. Bottom - the nuclear singlet state is transformed to magnetization of the protons using the magnetic inequivalence caused by non-symmetric coupling to the $^{13}\mathrm{C}$ site.}
    \label{fig:scheme}
\end{figure}

\begin{figure}[t]
    \centering
    \includegraphics[width=0.45\textwidth]{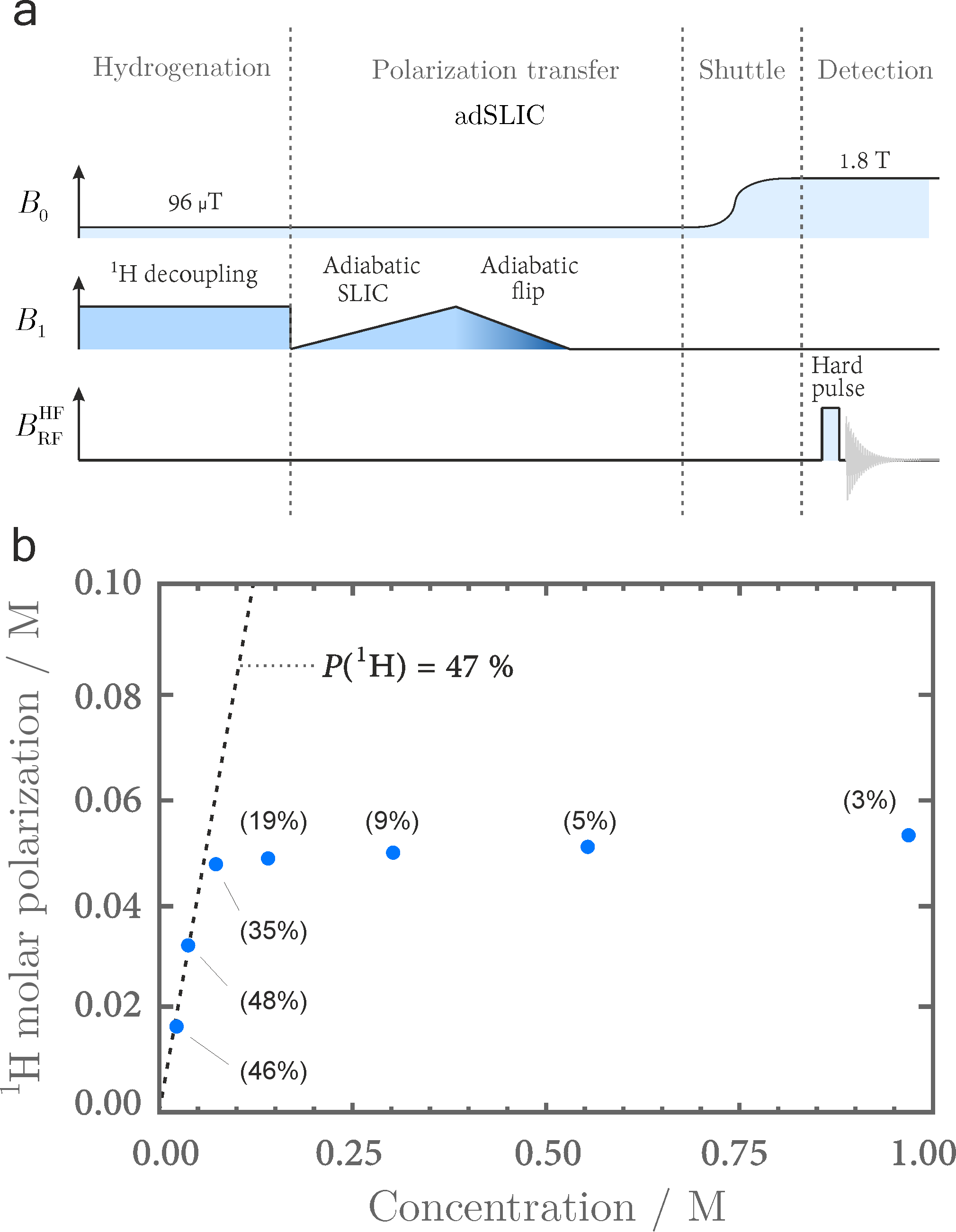}
    \caption{(a) The magnetic field sequence employed for the adSLIC polarization experiments. The procedure begins by hydrogenating a solution of \DMAD at 96~$\mu$T and under continuous wave irradiation at \Hone Larmor frequency. Polarization transfer is performed by ramping up the amplitude of an on-resonant rf field (adiabatic SLIC). The magnetization is rotated to $B_0$ by ramping down both the amplitude and frequency of the $B_1$ field, with the frequency shift which is depicted as color shading. The sample is then transported to a benchtop NMR magnet (indicated as high-field - HF) where signal is acquired after a hard rf pulse. (b) \Hone molar polarization of hyperpolarized target \DMM as a function of concentration achieved by the adSLIC sequence. The amplitude sweep duration was set to 2 seconds. The dashed line represents a fixed polarization level of $47\%$, and polarization levels are shown in parentheses next to the data points.}
    \label{fig:adSLIC}
\end{figure}

In this work we show that this phenomenon, previously associated with the excitation of multiple echoes and experimental artifacts, can be sufficiently strong to interfere with polarization-transfer sequences in hyperpolarized samples.
We obtained high \Hone molar polarization using the hydrogenation reaction of \DMAD with parahydrogen as shown in Fig.~\ref{fig:scheme}. This reaction produces \DMM in which the two \Hone spins from parahydrogen remain entangled in a nuclear singlet state, but are no longer magnetically equivalent due to different $J$-couplings to the \Cth site.
This inequivalence enables conversion of the singlet spin order into magnetization by ramping the amplitude of a transverse magnetic field oscillating in resonance with the \Hone nuclei as shown in Fig.~\ref{fig:adSLIC}(a).
%
This method is known as adiabatic Spin-Lock Induced Crossing (adSLIC) and it can induce complete conversion of singlet order to transverse \Hone magnetization~\cite{devience_preparation_2013,marshall_radio-frequency_2023,Pravdivtsev2016RobustCO,RODIN201814}. 

At low \DMM concentrations ($<$100\,mM) we consistently observe approximately 47\% \Hone polarization following the hyperpolarization process, a factor of $\sim$2 below the theoretical 100\% limit, presumably because of imperfect transfer from adSLIC and losses due to spin relaxation. However, if the product concentration is increased beyond this value, the corresponding increase in molar polarization becomes highly nonlinear, and reaches a limit at $\sim$$60\,$mM of \Hone molar polarization, as shown in Fig.\ref{fig:adSLIC}(b).
Constant molar polarization independent of product concentration means that in this regime the polarization is inversely proportional to the concentration of the polarized target.
We hypothesize that this limit is due to a large dipolar field 
that emerges during the transformation 
of the singlet state into observable magnetization, which disrupts the adSLIC polarization transfer step.
This is not a radiation damping effect, as the untuned and large excitation coil used for these low-field experiments couples too weakly to the nuclear spins to induce any appreciable radiation damping, and we have seen a similar limit is encountered using magnetic field cycling, a simpler polarization transfer method not requiring a transverse ($B_1$) field~\cite{johannesson2004transfer}. 
Our observation presents a substantial obstacle for achieving high molar polarization, and likely holds relevance for many other hyperpolarization techniques involving high sample concentrations or polarization, such as dynamic nuclear polarization or spin-exchange optical pumping~\cite{eills_spin_2023}.


Here we propose a solution to overcome this 
challenge by implementing a Lee-Goldburg decoupling sequence which is commonly used in solid-state NMR to average out strong dipolar interactions~\cite{lee_nuclear-magnetic-resonance_1965,goldburg_nuclear_1963}. We explain how to combine this 
with suitable periodic modulation to re-establish a polarization transfer equivalent to 
adSLIC that we refer as LG-adSLIC. In our experimental
work we verify the principle and demonstrate that the application of this pulse sequence leads 
to an order of magnitude improvement over the previous limit, yielding up to $\sim$450\,mM \Hone molar  polarization. The achieved improvement is primarily limited by coil inhomogeneities 
in our device and can in principle be enhanced further. 
This should enable new PHIP
applications involving high molar polarization, and may help to mitigate distant dipolar field effects in other areas of hyperpolarized NMR.


{\em Dipolar fields and Lee-Goldburg decoupling--} 
Let us first consider the dipolar field generated in an ensemble of single spin-1/2 nuclei in the presence of off-resonant, Lee-Goldburg (LG) decoupling~\cite{lee_nuclear-magnetic-resonance_1965,goldburg_nuclear_1963}.
We may then easily extend our considerations to the case of a heteronuclear three-spin system incorporating polarization transfer during the said decoupling.

The Hamiltonian of an isolated single nuclear spin ensemble subject to external fields provided by magnetic coils and  internal dipolar fields generated by the spin ensemble can be written in three terms: 
\begin{align}
    H_{I}(t) &= H_{\mathrm{0},I}+H_{\mathrm{LG},I}(t) + H_{\mathrm{DF},I} \label{eq:H_I} \\ \nonumber
    &= - \gamma_I B_{0} I_z - \gamma_I B_\mathrm{LG}(t) I_x + H_{\mathrm{DF},I}, 
\end{align}
where $\gamma_I$ is the nuclear gyromagnetic ratio, $B_0$ is an external static magnetic field, $B_\mathrm{LG}(t)$ is an external transverse field and
$H_{\mathrm{DF},I}$ takes into account internal magnetic field flux component due to all dipolar field contributions from distant nuclear spins. 

Under most NMR conditions, the last term is negligible and can be ignored.
At high concentrations or large polarization levels, however, dipolar fields can significantly impact the 
system's dynamics as it scales with number of spins. 
This interaction between each spin and the sample is complex and may be described 
either microscopically, accounting for the dipolar interaction between all spins explicitly 
\cite{richter_imaging_1995}, or by adopting a mean-field description, which defines the dipolar field generated by a spatially homogeneous sample \cite{levitt_demagnetization_1996}. 
For our purposes, these two descriptions yield equivalent results, and we use the mean field approach. 

Expressing $H_{\mathrm{DF},I}$ in the frame rotating at frequency $\omega$ of the continuous-wave transverse field $B_\mathrm{LG}(t)$ and discarding rapidly oscillating terms gives the state-dependent Hamiltonian (cf.~Eq.~(16) from \cite{heIntermolecularMultipleQuantum1993})
\begin{equation}
    H'_{\mathrm{DF},I} = \Delta_\mathrm{DF} \left[\langle \vec{I}\rangle \cdot \vec{I}  
    - 3\langle I_z\rangle I_z\right]\, \label{eq:H_dipI'}
\end{equation}
where $\langle {\bf I}\rangle = \overline{\langle\psi(t)|{\bf I}|\psi(t)\rangle}$ and the over-bar indicates an average over the spin ensemble.  
Assuming a spatially homogeneous sample, we have
\begin{equation}
    \Delta_\mathrm{DF}=\Delta_\mathrm{DF}({\vec r}_l) = \sum_{k\neq l} \frac{\mu_0 \gamma_I^2}{4\pi} \frac{1 - 3({\vec e}_z{\vec r}_{kl})^2/|{\vec r}_{kl}|^2}{|{\vec r}_{kl}|^3} \label{eq:Delta_dip}
\end{equation}
with ${\vec r}_{kl} = {\vec r}_k - {\vec r}_l$ where ${\vec r}_l$ denotes the position of nucleus $l$. 
While in the general case $\Delta_\mathrm{DF}({\vec r}_l)$ depends on the position of spin $l$ and of all the other 
molecules and their diffusive motion in the sample relative to spin $l$, the general structure of Eq.~\ref{eq:H_dipI'} 
remains independent of it with time dependence suppressed as well. The contribution to $\Delta_\mathrm{DF}({\vec r}_l)$ from nearby spins is suppressed by molecular
diffusion because Eq.~\ref{eq:Delta_dip} vanishes when $\vec{r}_{kl}$ is averaged over a spherically symmetric 
volume~\cite{jeener_dynamical_1999,jeener_dynamical_2002,warren_generation_1993}. Hence, only distant nuclei contribute to the dipolar field.

In order to minimize the influence of the dipolar field we make the $B_\mathrm{LG}$ field off-resonant with respect to the Larmor frequency such that:
\begin{align}\label{eq:B_RF}
    -\gamma_I B_\mathrm{LG}(t) &= 2\omega_\mathrm{LG} \sin\theta \cos(\omega t), \\ \nonumber
    \omega &=\omega_{0,I} - \omega_\mathrm{LG} \cos\theta\, ,
\end{align}
where $\omega_{0,I}$ is the Larmor frequency of spin $I$ and factor of 2 takes into account the average power of the linearly oscillating transverse field.
The total Hamiltonian $H_I$ in the rotating frame then becomes:
\begin{align}
    H'_{I} &=  H'_{1,I}+H'_{\mathrm{DF},I} \\&=\omega_\mathrm{LG} (\cos\theta I_z + \sin\theta I_x) + H'_{\mathrm{DF},I}\nonumber, 
\end{align}
where $\omega_{LG}$ and $\theta$ define amplitude and orientation of a new effective 
field. The eigenbasis of $H'_{1,I}$ 
leads to the tilted operators
\begin{align}
 \tilde{I}_x &= -\sin\theta I_z + \cos\theta I_x\, , \nonumber \\ 
 \tilde{I}_y &= I_y\, ,  \label{tildeoperators}\\
 \tilde{I}_z &=  \cos\theta I_z + \sin\theta I_x\, . \nonumber
\end{align}
Rewriting the Hamiltonian in this basis, moving to a second interaction frame of $H'_{1,I}$ establishes what we
henceforth refer to as the effective field frame. Neglecting rapidly oscillating terms, we find that dipolar field Hamiltonian in this frame becomes
\begin{align}
    H''_{\mathrm{DF},I}(\theta) &= \Delta_\mathrm{DF}\frac{(3\cos^2\theta -1)}{2}\left[ \langle \tilde{\vec{I}}\rangle \tilde{\vec{I}} - 3 \langle \tilde{I}_{z}\rangle \tilde{I}_{z}\right], \label{eq:H_DFdp}
\end{align}
which vanishes at the magic angle $\theta_\mathrm{M} = \arccos\sqrt{1/3}$.
Note that at the magic angle is exactly Lee-Goldurg decoupling condition which is used to miminize the effects of dipolar coupling.

{\em Polarization transfer in the effective field frame--} 
An extension of our off-resonant decoupling to singlet-to-magnetization transfer to achieve \Hone magnetization in \DMM (Fig.~\ref{fig:scheme}) may be given as follows.
First, the total Hamiltonian of coupled heteronuclear 3-spin system may be given by extending Eq.~\ref{eq:H_I} to a modified form:
\begin{equation}
    H(t) = H_\mathrm{spin} + H_\mathrm{rf}(t) + H_{\mathrm{DF}}.
    \label{Hamiltonian}
\end{equation} 
The dipolar field Hamiltonian $H_\mathrm{DF}$ inherits the same structure as $H_{\mathrm{DF},I}$ by using substitution $I_i\rightarrow I_i^\Sigma$ with $I_i^{\Sigma} := I_{1,i}+I_{2,i}$. Note that we do not consider corresponding terms from the $S$ spins as these remain unpolarized throughout the experiment while merely experiencing a Zeeman shift from the $I$-induced dipolar field, and this shift does not contribute to the dynamics of the $I$ spins.
In the present case $I$ and $S$ spins are \Hone and \Cth nuclei, respectively.
Here, $H_\mathrm{spin}$ now includes Zeeman interaction for all spins $I$ and $S$ as well as $J$-couplings between them:
\begin{align}\label{eq:H_spin2}
    H_\mathrm{spin} &=  H_\mathrm{0} + H_\mathrm{J}^{II}+H_\mathrm{J}^{IS} \, ,\\
    H_\mathrm{0} &= -\gamma_I B_0 I^\Sigma_z - \gamma_S B_0 S_z \, ,\nonumber\\
    H_\mathrm{J}^\mathrm{II} &= 2\pi J \vec{I}_1\cdot \vec{I}_2\, , \nonumber \\
    H_\mathrm{J}^\mathrm{IS} &= 2\pi(J_1 I_{1,z}+ J_2 I_{2,z}) S_z.\nonumber
\end{align}

Hamiltonian $H_\mathrm{rf}$ describes the external transverse fields that are applied to $I$ spins and is given by
\begin{equation}
    H_\mathrm{rf}(t) = -\gamma_I(B_\mathrm{LG}(t) + B_\mathrm{mod}(t)) \cdot I^\Sigma_x.
\end{equation}
The transverse field is now decomposed into two terms. The first term is the LG decoupling field $B_\mathrm{LG}(t)$ as written in Eq.~\ref{eq:B_RF}
and is used to mitigate dipolar field by selecting appropriate effective field angle.
The singlet-to-magnetization transfer using adSLIC is performed during the said decoupling.
Therefore, a second and lesser component $B_\mathrm{mod}$ is applied which slightly modulates 
the decoupling field. The modulation field is given by
\begin{align}
    -\gamma_I B_\mathrm{mod}(t) = -2\sin(\omega t)\cdot 2\,\omega_2(t) \cos(\omega_\mathrm{mod} t+\phi),
    \label{eq:Bmod}
\end{align}
where $\omega_\mathrm{mod}$ is the modulation frequency and the time-dependent amplitude $\omega_2(t)$ is needed for adiabatic polarization transfer. The second factor of 2 is added to further compensate for linear polarization of the applied field.

Combining the terms and expressing the total Hamiltonian (Eq.~\ref{Hamiltonian}) in the Zeeman interaction frame we obtain
\begin{eqnarray}
    H'(t) &=& H_\mathrm{J}^{II}+H_\mathrm{J}^{IS}+H'_{\mathrm{DF}}  \nonumber\\  
    && + \omega_\mathrm{LG} (\cos\theta I^{\Sigma}_z+\sin\theta I^{\Sigma}_x)\nonumber\\
    &&  + 2\omega_{2}(t)\cos(\omega_\mathrm{mod} t+\phi) I^{\Sigma}_y
\end{eqnarray}
which simplifies with tilted operators in Eq.\ref{tildeoperators} to 
\begin{eqnarray}
    H'(t) &=& H_\mathrm{J}^{II}+H_\mathrm{J}^{IS}+H'_{\mathrm{DF}}  \nonumber\\  
    && +\omega_\mathrm{LG}\tilde{I}^{\Sigma}_z\nonumber\\
    &&  + 2\omega_{2}(t)\cos(\omega_\mathrm{mod} t+\phi) \tilde{I}^{\Sigma}_y.
\end{eqnarray}

It is evident that the last two terms in Hamiltonian mimic the case of $I$ spins being exposed to a static  field of amplitude $\omega_\mathrm{LG}$ and an oscillating transverse field with amplitude $2\omega_2$.
Therefore, if the modulating field is in resonance with the effective field such that $\omega_\mathrm{mod}=\omega_\mathrm{LG}$ we can further simplify the Hamiltonian by expressing it in the doubly-rotating frame and discarding rapidly oscilating terms:
\begin{eqnarray}
\label{eq:eff_pulse}
	H''(\theta,t) &= \tilde{H}_\mathrm{J}^{II}+\cos\theta \tilde{H}_\mathrm{J}^{IS} + H''_\mathrm{DF}(\theta)\\ &+ \omega_{2}(t) ( \cos\phi\tilde{I}^{\Sigma}_y-\sin\phi\tilde{I}^{\Sigma}_x)\nonumber
\end{eqnarray}
where we find that heteronuclear $J$-coupling Hamiltonian is scaled by the cosine of effective angle.
The tilde indicates the use of tilted operators retaining the structure of Eq.\ref{eq:H_spin2}.
For $\theta \neq \theta_\mathrm{M}$, the dipolar coupling $H''_\mathrm{DF}$ is partially suppressed compared to the original $H_{\mathrm{DF}}$ (cf.~Eq.~\ref{eq:H_dipI'}) whereas at the magic angle we get $H''_\mathrm{DF} = 0$ 
and recover the dipolar-field-free Hamiltonian where at phase $\phi=0$ it leads to:
\begin{align}
	H''_{\theta_\mathrm{M}}(t) &= \omega_{2}(t)\tilde{I}^{\Sigma}_y +\tilde{H}_\mathrm{J}^{II}+\frac{1}{\sqrt{3}} \tilde{H}_\mathrm{J}^{IS} .
\end{align}
 
As a result of LG decoupling, the adSLIC sequence achieving magnetization on $I$ spins (\Hone in the present case) can be implemented in the effective field frame or exactly at LG frame via $B_\mathrm{mod}(t)$ without obstruction by dipolar fields. As the derivation relies on the scale hierarchy $\omega_0\gg\omega_\mathrm{LG} \gg \Delta_\mathrm{DF}, \omega_2$, we use an adiabatic SLIC \cite{devience_preparation_2013, marshall_radio-frequency_2023,Pravdivtsev2016RobustCO,RODIN201814} to achieve robust transfer. It is important to stress that while the level anti-crossing condition for SLIC does not change ($\omega_2=2\pi J$) the transfer rate and thus adiabaticity is scaled by $1/\sqrt{3}$ as a consequence of tilted effective field.
This approach is also suited for implementing other homonuclear NMR sequences by selecting phase and time-dependent amplitude in Eq.\ref{eq:eff_pulse}. 

\begin{figure}[h]
    \centering
    \includegraphics[width=0.45\textwidth]{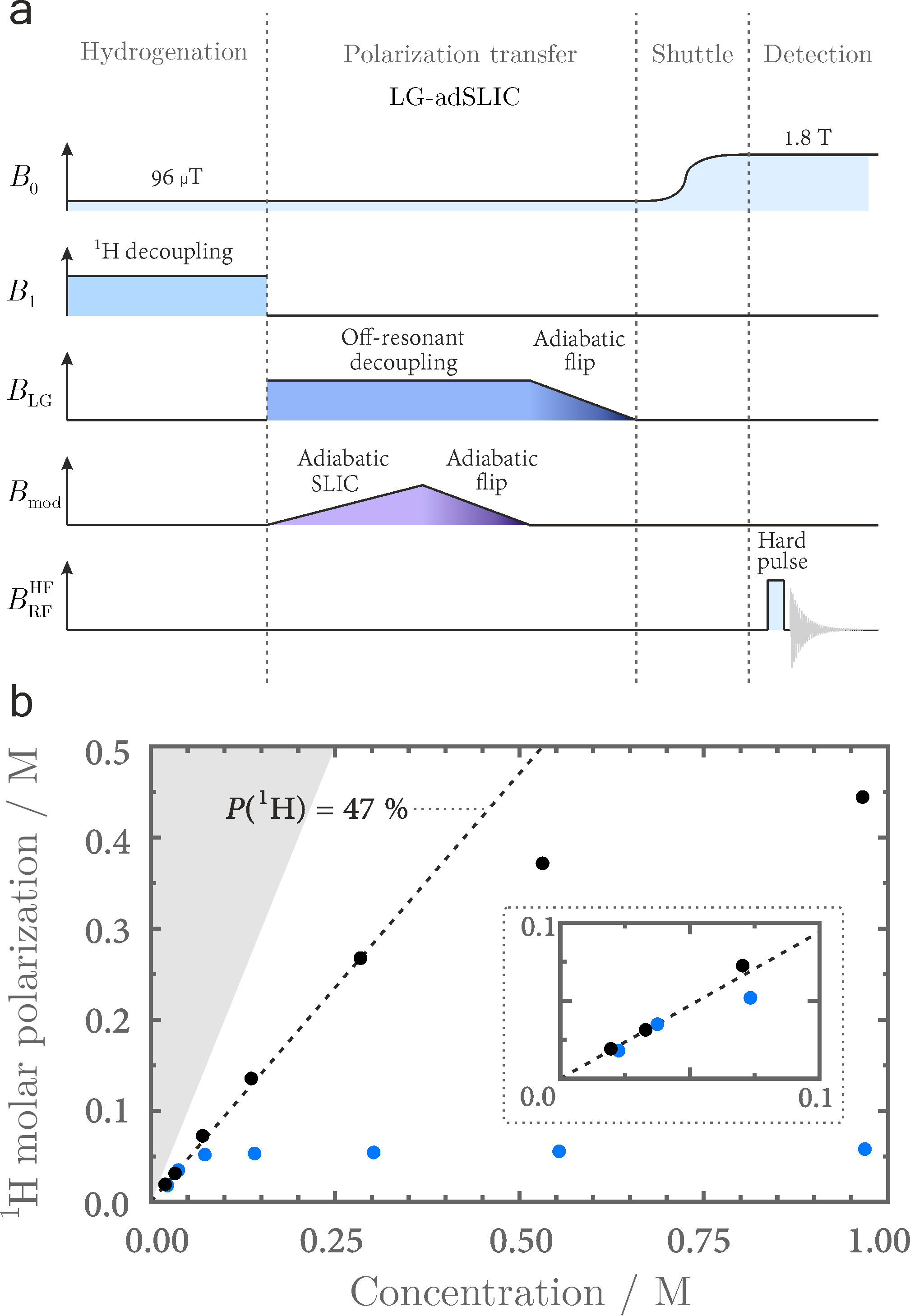}
    \caption{(a) A modified magnetic field sequence LG-adSLIC that includes Lee-Goldburg decoupling. Polarization transfer is performed with a modulation field (Eq.~\ref{eq:Bmod}) mimicking the adSLIC transfer in Fig.\ref{fig:adSLIC}(a) while under strong continuous irradiation with a resonance shift. To rotate the magnetization to align with $B_0$, first the amplitude and frequency of the $B_\mathrm{mod}$ field was ramped down to rotate the magnetization along the effective field, and then the amplitude and frequency of the $B_\mathrm{LG}$ field were ramped down to rotate the magnetization along $B_0$.
    (b) \Hone molar polarization of hyperpolarized target \DMM as a function of concentration achieved by the LG-adSLIC sequence (black dots) compared to the previous results when LG decoupling was omitted (blue dots). The amplitude sweep duration was set to 2\,s in both cases and the Lee-Goldburg effective field amplitude was set to $\omega_\mathrm{LG}=2\pi\,600\,$Hz 
    (more details in Methods). The shaded area indicates the nonphysical region in which \Hone polarization exceeds 100\%. 
    A scaled inset plot is provided for clarity.}
    \label{fig:LGadSLIC}
\end{figure}

\begin{figure}[t]
    \centering
    \includegraphics[width=0.45\textwidth]{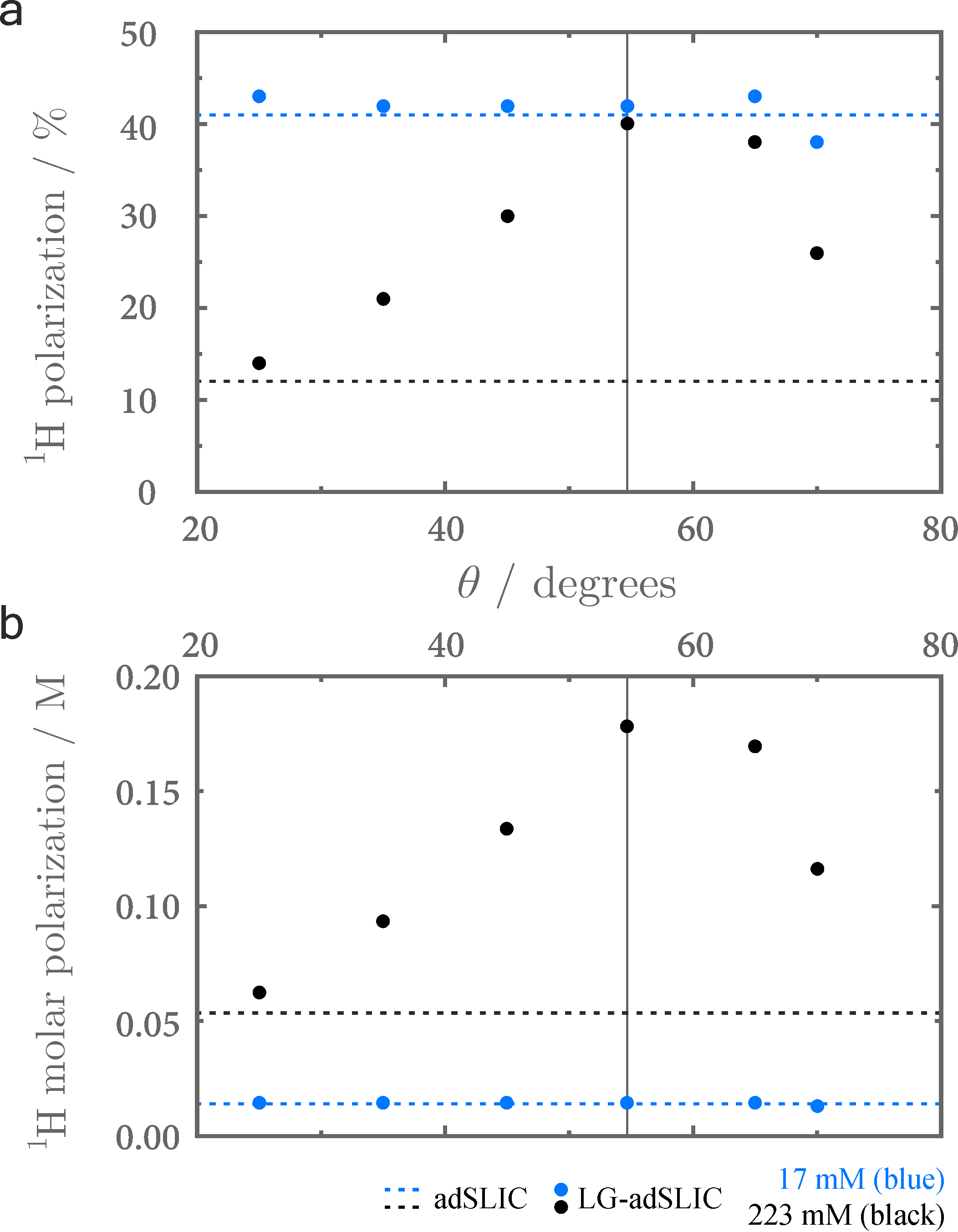}
    \caption{(a) \Hone spin polarization and (b) \Hone molar polarization of hyperpolarized \DMM as a function of effective angle $\theta$ used in the LG-adSLIC sequence (Fig.~\ref{fig:LGadSLIC}(a)). Data points acquired at \DMM concentrations of  $17\,$mM and $223\,$mM  are shown in blue and black, respectively. The amplitude sweep duration was set to 4\,s and the effective field amplitude was set to $\omega_{LG}=2\pi\,400\,$Hz (see Methods for more details). Dashed lines indicates the level of polarization acquired with the adSLIC sequence in Fig.~\ref{fig:adSLIC}(a) at high and low product concentrations. The magic angle value is shown as a vertical line.}
    \label{fig:AnglePlot}
\end{figure}
{\em Methods--}
The precursor solution for \DMM was prepared by dissolving $5\,$mM \Cat catalyst (CAS number: 79255-71-3) into acetone-d$_6$. For the experiments with varied \DMM concentrations, precursor concentrations were prepared in this order:  20, 40, 80, 160, 320, 640, $1080\,$mM.
Two precursors concentration were used in Fig.~\ref{fig:AnglePlot}, $20\,$mM and $300\,$mM for the blue and black points in, respectively.

Parahydrogen was produced by ARS parahydrogen generator packed with an iron monohydrate catalyst, running at 22~K temperature and producing gas with a para-enrichment level of $\sim 93\,$\%. 

Figure~\ref{fig:adSLIC} and 
Each experiment starts by injecting $500\,\mu$L of solution into a tube and bubbling para-enriched hydrogen gas through the solution at $10\,$bar pressure at a bias field of $96\,\mathrm{\mu}$T. This is followed by nitrogen bubbling at $10\,$bar to stop the reaction proceeding further. To avoid fast singlet order decay, resonant \Hone decoupling is provided throughout the entire bubbling period which in all experiments was fixed to $30$ seconds~\cite{marshall_radio-frequency_2023,Dagys_deuteron_2022}.

Polarization transfer was performed following two different protocols as displayed in Fig.~\ref{fig:adSLIC}(a) and Fig.~\ref{fig:LGadSLIC}(a).
The first one consisted of a transverse field swept up from 0 to $(2\pi)25$~Hz in amplitude (with respect to \Hone), followed by an adiabatic flip pulse. The flip pulse was arranged by ramping the transverse field amplitude down in $1\,$second with gradual carrier frequency shift ($\omega_0+\Delta\omega_0$) of $\Delta\omega_0=-(2\pi)200\,$~Hz. No decoupling was applied during the polarization transfer.

The second method included an off-resonant (Lee-Goldburg) decoupling (cf.~Eq.~\ref{eq:B_RF}) during the polarization transfer to minimize the influence of the dipolar field. The effective field amplitude $\omega_\mathrm{LG}$ was set to $(2\pi)\,600\,$Hz and $(2\pi)\,400$Hz for experiments in Fig.~\ref{fig:LGadSLIC} and Fig.~\ref{fig:AnglePlot}, respectively.
After the polarization transfer, a flip pulse was performed by ramping the transverse field amplitude down in $1\,$second with a gradual decoupling field frequency shift ($\omega+\Delta\omega$) of $\Delta\omega=-(2\pi)200\,$~Hz.
Polarization transfer during LG decoupling was initiated by ramping modulation field (Eq.\ref{eq:Bmod}) amplitude from 0 to $(2\pi)25\,$Hz (with respect to \Hone). The modulation frequency was set to match the effective field amplitude ($\omega_\mathrm{mod}=\omega_\mathrm{LG}$). To perform adiabatic pulse to flip magnetization along the effective field, the modulation amplitude was ramped down in $1\,$second with gradual modulation frequency shift ($\omega_\mathrm{mod}+\Delta\omega_\mathrm{mod}$) of $\Delta\omega_\mathrm{mod}=-(2\pi)200\,$~Hz.

The \Hone free-induction decays were excited by a small flip angle pulse of $(2\pi)20\,$kHz rf amplitude and recorded with 131~k point density at a spectral width of $400\,$ppm. Additional \Hone decoupling was used for all experiments. Thermal equilibrium \Hone spectra were recorded at room temperature with a recycle delay of $90\,$s and with a $90$ degrees flip angle pulse. 
Polarization levels were calculated by comparing the \Hone signals of hyperpolarized and thermally polarized samples.
When estimating polarization level, the scaling factor of different excitation pulses was taken into account.
The concentration of \DMM was determined by comparing the thermal equilibrium signal to the signal of an external standard of known-concentration measured under the same conditions.
The molar polarization was calculated as the product of the concentration, the spin-polarization, and the number of \Hone sites in the molecule (two in the present case).

{\em Results -- }
Implementing LG decoupling into the polarization process leads to a significant improvement in the molar polarization that can be obtained at high sample concentrations. The experimental sequence and results are shown in Fig.\,\ref{fig:LGadSLIC}. 
Operating under the same experimental conditions and getting two contrasting outcomes using adSLIC and LG-adSLIC is a strong indication that the limited molar polarization is not related to chemical impurities disrupting the polarization process. 
The linear scaling of molar polarization with product concentration (as indicated by the dashed line in Fig.~\ref{fig:LGadSLIC}) persists to higher concentration values when LG decoupling is used.
There is still a decrease in sample polarization at molar polarizations above $\sim$300\,mM, and we attribute this to insufficient LG decoupling at such high sample magnetization. In principle this could be remedied by employing a stronger LG decoupling field, but this additionally requires higher $B_1$ field homogeneity which was impractical to implement on our equipment.

The efficacy of LG decoupling on \Hone polarization is investigated further by varying the effective field angle $\theta$, and the results are shown in Fig.~\ref{fig:AnglePlot}.
At low concentration of \DMM (17\,mM) no dependence on the angle $\theta$ was observed as the sample dipolar field is negligible and so the LG decoupling does not affect the polarization. This was not the case at higher concentration of \DMM (223\,mM) where LG decoupling is important for obtaining high polarization.
The maximum polarization was achieved when setting the effective angle to the magic angle $\theta=\theta_\mathrm{M}$ which is consistent with prediction from Eq.~\ref{eq:H_DFdp}. 
We reiterate that radiation damping is not expected to play a role in these experiments as the sample-coil coupling is negligible since low excitation frequencies were used and the large coil volumes result in a low filling factor.


{\em Conclusions --}
In this work we observe that the achievable molar polarization in PHIP-polarized samples is limited to approximately 60~mM, independent of the  product concentration above a threshold of approximately 100~mM. 
This limit was observed in samples of \DMM following the application of a low-field adiabatic spin-lock induced crossing (adSLIC) sequence to induce \Hone singlet-to-magnetization conversion. 
Our findings suggest the limited molar polarization is due to a distant dipolar field originating from the polarized \Hone spins as the sample becomes magnetized. 
The internal magnetic field along the cylinder axis in a sample of \Hone spins at 60~mM molar polarization is approximately 214~nT which would contribute 9~Hz to the Zeeman interaction.
This value is comparable to the amplitude of the transverse field used and spin-spin couplings in the molecule and thus disrupts the adSLIC pulse sequence effectively. 
We have seen that a similar limit is encountered using simpler singlet-to-magnetization sequences such as adiabatic magnetic field cycling (MFC) as the bias field inducing the polarization transfer is in sub-microtesla regime as well.

To negate this adverse effect we implemented Lee-Goldburg decoupling, leading to an improvement in the achievable molar polarization by an order of magnitude. 
Our work highlights that further improvements in hyperpolarization can lead to circumstances where NMR pulse sequences can be disrupted by high internal sample magnetization and could complicate interpretation.
Sequences which incorporate averaging of the dipolar interaction can help to reduce and diagnose this phenomenon.
This is crucial, as hyperpolarization methods that produce highly-polarized solutions have become increasingly prevalent in recent years.

{\em Acknowledgements --} The authors acknowledge financial support by the German Federal Ministry of Education and Research
(BMBF) under the funding program quantum technologies - from basic research to market via the project QuE-MRT (FKZ: 13N16447) as well as the EIC Transition project MagSense (grant no. 101113079). 
We acknowledge support received from EPSRC, UK by the grants EP/V055593/1 and EP/W020343/1.
This project has received funding from the European Union’s Horizon 2020 Research and Innovation Programme under the Marie Sk\l{}odowska-Curie Grant Agreement 101063517.
MBP and MK additionally acknowledge financial support by the ERC Synergy grant HyperQ (grant no. 856432).

\newpage

{\em Appendix}
\renewcommand{\thefigure}{A\arabic{figure}}
\setcounter{figure}{0}
\begin{figure}[H]
    \centering
    \includegraphics[width=0.45\textwidth]{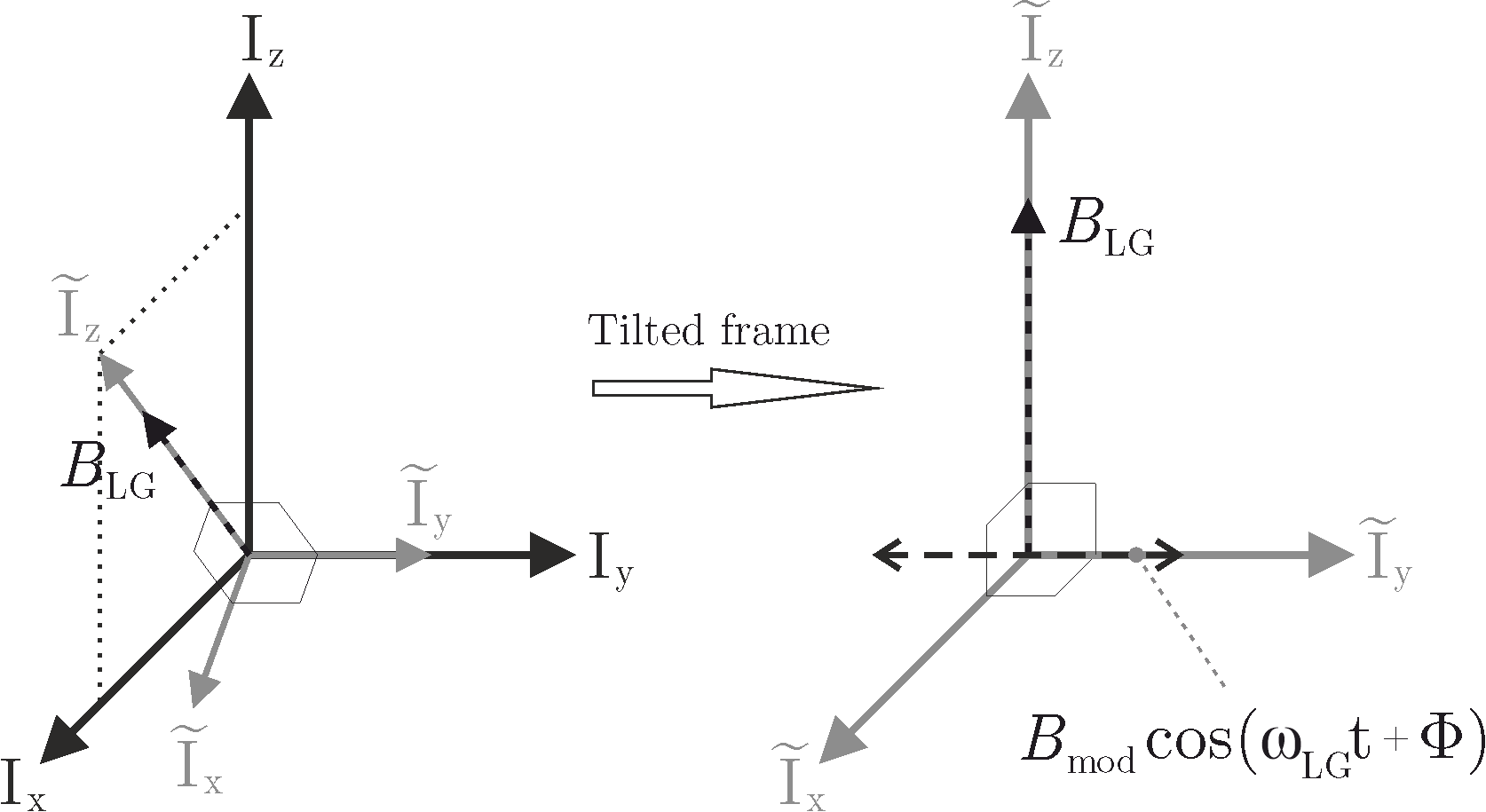}
    \caption{The vector representation of Lee-Goldburg (LG) frame. On the left - consider a frame rotating with Larmor frequency of spin $I$. The off-resonant nature of Lee-Goldburg decoupling leads to a new field $B_\mathrm{LG}$ tilted in XZ-plane with transverse component representing the amplitude of decoupling and longitudinal component reflecting resonance mismatch. On the right - the original rotating frame is now tilted-back such that $B_\mathrm{LG}$ is a principal axis in a so-called LG frame. To set up a pulse in this frame, additional field $B_\mathrm{mod}$ oscillating with frequency $\omega_\mathrm{LG}=-\gamma_\mathrm{I}B_\mathrm{LG}$ needs to be applied. Such component along $I_y$ (note that $\tilde{I}_y = I_y$) can be generated by phase-shifting the LG decoupling by $3\pi/2$ which becomes a sine-wave as shown in equation~\ref{eq:Bmod}.}
    \label{fig:vector_pic}
\end{figure}

\bibliography{LGPHIP.bib} 

\end{document}